\documentclass[11pt, letterpaper]{article}

\pdfoutput=1

\usepackage{amsmath}
\usepackage{amsfonts}
\usepackage{amssymb}
\usepackage[colorlinks,citecolor=blue,urlcolor=blue,bookmarks=false,hypertexnames=true]{hyperref}
\usepackage{cite}

\usepackage{graphicx, physics, subcaption}
\usepackage{verbatim }

\numberwithin{equation}{section}

\usepackage{color}

\usepackage{amsmath, amssymb, graphics, epsfig, graphicx}
\usepackage{epsf}
\usepackage{epstopdf}
\usepackage {amssymb}
\newcommand{\nc}{\newcommand}
\nc{\ba}{\begin{eqnarray}}
\nc{\ea}{\end{eqnarray}}

\newcommand{\calR}{{\cal{R}}}
\newcommand{\calP}{{\cal{P}}}

\def\bfk{{\bf k}}

\def\bfx{{\bf x}}

\nc{\cN}{ {\cal{N}} }

\newcommand{\CMB}{\text{\tiny CMB}}

\begin{document}

\def\thefootnote{\fnsymbol{footnote}}

\begin{center}

{\bf \large Induced Gravitational Waves from Ultra Slow-Roll Inflation\\ and Pulsar Timing Arrays Observations
}
\\[0.5cm]

{ Hassan Firouzjahi,  \footnote{firouz@ipm.ir}
Alireza Talebian \footnote{talebian@ipm.ir}  }
\\[0.5cm]
 
 {\small \textit{$^1$School of Astronomy, Institute for Research in Fundamental Sciences (IPM) \\ P.~O.~Box 19395-5531, Tehran, Iran
}}\\

\end{center}

\vspace{.3cm}
\hrule
\begin{abstract}

The stochastic gravitational wave background (SGWB) detected recently by the pulsar timing arrays (PTAs) observations may have cosmological origins. In this work we consider a model of single field inflation containing an intermediate phase of ultra slow-roll. Fixing the amplitude of the peak of curvature perturbations  by the PBHs bounds we calculate the  gravitational waves (GWs) induced from the  curvature perturbations enhanced during USR.  The spectrum of the induced GWs depends on the sharpness of the transition from the USR phase to the final attractor phase as well as to the duration of the USR period.  While the model can accommodate  the current PTAs data but it  has non-trivial predictions  for the induced GWs on higher frequency ranges  which can be tested by future observations. 

\end{abstract}

\vspace{0.3cm}

\newpage

\section{Introduction}
\label{Introduction}

 There are indications of detection of stochastic gravitational waves background (SGWB) from recent pulsar timing arrays (PTAs) around the frequency 
 range  $ \sim 10 ~\mathrm{nHz}$ as reported in  NANOGrav~\cite{NANOGrav:2023gor}, Parkers PTA~\cite{Reardon:2023gzh}, European PTA~\cite{Antoniadis:2023ott} and the China PTA~\cite{Xu:2023wog}.  
 These signals may have cosmological origins as well as astrophysical interpretations. A natural  astrophysical interpretation of the observed SGWB is the superpositions of  gravitational waves (GWs)  signals from the merging of binary supermassive black holes \cite{NANOGrav:2023gor}.  On the other hand, if the observed signal has cosmological origins, this can open a new window into observing the primordial universe and to probe physics beyond the Standard Model of particle physics. Among possible cosmological interpretations of the SGWB
are the GWs induced from the enhanced scalar perturbations on small scales generated during inflation, first order cosmological phase transitions \cite{Kosowsky:1992rz, Kamionkowski:1993fg, Caprini:2007xq},  domain walls or cosmic strings \cite{Hindmarsh:2013xza, Kibble:1976sj,Vilenkin:1981bx, Caldwell:1991jj,Vilenkin:1981zs}, see \cite{NANOGrav:2023hvm, Antoniadis:2023zhi} for further review. It should be noted that  the  previous  NANOGrav 12.5 year data \cite{NANOGrav:2020bcs} also indicated traces of  SGWB with a flat spectrum in a narrow range of $\mathrm{nHz}$ frequency which initiated interests to look for the origins of this signal \cite{Vagnozzi:2020gtf,Zhou:2020kkf,Domenech:2020ers,Dandoy:2023jot}.

Scalar induced gravitational waves (SIGW) from the enhanced  scalar perturbations on small scale  during inflation \cite{Ananda:2006af, Baumann:2007zm, Bugaev:2009zh, Assadullahi:2009nf, Alabidi:2012ex, Cai:2018dig, Pi:2020otn, Balaji:2022dbi, Talebian:2022cwk, Domenech:2021ztg}  is a mechanism which can explain the observed SGWBs  \cite{NANOGrav:2023hvm, Antoniadis:2023zhi}. 
In this mechanism, the GWs are sourced at the second order in perturbation theory via their interaction with the scalar sector generated during inflation. Typically, this setup requires that the amplitude of scalar perturbations to grow by about seven orders of magnitude compared to the observed CMB scale. Consequently, this mechanism can yield to 
primordial black holes (PBHs) formation which may comprise all or parts of dark matter energy density \cite{ Carr:2016drx, Carr:2020xqk, Sasaki:2018dmp, Ozsoy:2023ryl, Byrnes:2021jka}. Naively speaking, suppose  $k_{\rm p}$ is the peak scale of perturbations responsible for PBH formation. Then PBH with a given mass have an induced GW counterpart with a peak frequency at around
	\begin{align}
		\label{fp}
		\dfrac{f_{\rm p}}{\rm 10nHz} \simeq 1.54 \times 10^{-7}\bigg(\dfrac{k_{\rm p} }{{\rm Mpc}^{-1}}\bigg) \,.
	\end{align}
	For example, in the case of PTAs signal, the amplified mode $k_{\rm p}$ leaves the horizon during inflation about $\sim 16$ e-fold after the CMB scale with $k_{\rm CMB} = 0.05~{\rm Mpc}^{-1}$ has crossed the horizon.

The setup of ultra slow-roll (USR) inflation  has been employed as a model  in which the primordial power spectrum can be enhanced to 
induce large SIGWs and PBHs  \cite{Ivanov:1994pa, Garcia-Bellido:2017mdw, Biagetti:2018pjj, Ragavendra:2020sop, Di:2017ndc, Liu:2020oqe, Hooshangi:2022lao, Hooshangi:2023kss, Ghoshal:2023wri}, for a review see \cite{Ozsoy:2023ryl, Byrnes:2021jka}. The USR setup is a single field model of inflation in which the potential is flat 
\cite{Kinney:2005vj, Morse:2018kda, Lin:2019fcz} and the inflaton velocity falls off exponentially so the curvature perturbations grow on superhorizon scales \cite{Namjoo:2012aa}. Since the  curvature perturbations grow on superhorizon scales the USR setup  violates  the Maldacena
non-Gaussianity consistency condition \cite{Maldacena:2002vr, Creminelli:2004yq} in single field models \cite{Namjoo:2012aa, Martin:2012pe, Chen:2013aj, Chen:2013eea, Akhshik:2015rwa, Mooij:2015yka, Bravo:2017wyw, Finelli:2017fml, Pi:2022ysn}. Originally, it was shown in \cite{Namjoo:2012aa} that the amplitude of local-type non-Gaussianity in USR setup is $f_{NL}=\frac{5}{2}$. This question was further examined in \cite{Cai:2018dkf} in which it was shown that the amplitude of $f_{NL}$ crucially depends on the sharpness of the transition from the USR phase to the final slow-roll (SR) phase. In an extreme sharp transition from the USR phase to the SR phase, which was assumed in \cite{Namjoo:2012aa}, $f_{NL}$ reaches its maximum value $\frac{5}{2}$. However, if the transition to the final stage is mild then the curvature perturbations evolve after the USR phase before it reaches to its final attractor value. Correspondingly, in a mild transition, 
the amplitude of $f_{NL}$ is washed out 
in the subsequent evolution and it ends up with a value at the order of the slow-roll parameters.  

Another important point  is the issue of loop corrections in this setup. 
This question was studied in various recent works \cite{yokoyama, Kristiano:2023scm, Riotto:2023hoz, Riotto:2023gpm, Choudhury:2023jlt, Choudhury:2023rks, Firouzjahi:2023aum, hr, Motohashi:2023syh, f, t, Firouzjahi:2023btw, Fumagalli:2023hpa}. Originally, it was argued in  \cite{yokoyama}, see also \cite{Kristiano:2023scm}, that loop corrections from small scale modes which leave the horizon during the USR phase can induce large corrections on  CMB scale modes. This was criticized in \cite{Riotto:2023hoz, Riotto:2023gpm} arguing, among other things,  
that for a mild transition the loop corrections will be less significant and the standard PBHs formation within the single field USR scenario is applicable. This question was studied in some details in \cite{Firouzjahi:2023aum} with emphasis on the effects of the 
sharpness of the transition from the intermediate USR phase to the final attractor SR phase. It was shown in \cite{Firouzjahi:2023aum} that for an arbitrarily sharp transition 
 the one-loop corrections can be very large, in line with  the results advocated in \cite{yokoyama, Kristiano:2023scm}. However, it was speculated in \cite{Firouzjahi:2023aum} that for a mild transition, the dangerous one-loop corrections are washed out during the subsequent evolution of the mode function after the USR phase. This conclusion was further examined in \cite{hr} confirming the physical expectations of \cite{Riotto:2023hoz, Riotto:2023gpm}. In summary, in order for the one-loop corrections on CMB scale modes to be harmless one needs a mild enough transition from the USR phase to the final attractor phase.

In this paper we employ the USR setup as a working mechanism to generate large 
SIGW as a possible explanations for the observed SGWB in the NANOGrav data \cite{NANOGrav:2023gor}. For various recent works on SIGWs as 
an explanation of the the PTAs data see  \cite{Vagnozzi:2023lwo,  Franciolini:2023pbf, Cai:2023dls, Inomata:2023zup, Wang:2023ost, Liu:2023ymk, Yi:2023mbm, Figueroa:2023zhu, Gu:2023mmd, Ebadi:2023xhq, Madge:2023cak}.

\section{The Setup}

 The setup which we use to enhance the primordial power spectrum 
 to induce  large GWs at the range of scales observed by the PTA observations  contains an intermediate phase of USR in single field inflation. 
We have a three-stage model of inflation in which the large CMB scale leaves the horizon at the early stage of inflation. The first stage of inflation proceeds say in about 16 e-folds or so. Then the USR phase takes over in which the potential is very flat and the curvature perturbations experience a growth on superhorizon scales. In order for 
 the curvature power spectrum to be under perturbative controls the USR phase has to be terminated followed by a final
  phase of SR inflation. A realistic setup requires that the transition from the first SR to the USR stage and from the USR phase to the final SR phase to be smooth.
  However, in order to follow the dynamics analytically, we consider an idealized situation in which the transition from the SR to USR and then to final SR phase are instantaneous. Assuming that USR phase is extended during the time interval 
  $ t_i \leq t \leq t_e$, we assume that the transitions at the starting point $t=t_i$ and at the final point $t=t_e$ to be instantaneous. While the transition to the final SR phase is instantaneous, but it will take time for the system to relax to its final attractor phase. We control this relaxation period by a sharpness parameter which plays important role in our analysis below. It it important  that the instantaneous gluing of the solutions should not be confused with the sharpness of the transition to the final attractor solution. 
 
 With the above discussions in mind, let us elaborate on the dynamics of our setup.
 During the first stage of inflation, $t<t_i$, the system follows an attractor phase and the dynamics of the inflaton field $\phi$ is given by the usual SR dynamics. The Hubble expansion rate $H\equiv \frac{\dot a}{a}$ is nearly constant in which $a(t)$
 is the FLRW scale factor. The small evolution of $H$ during the first stage of inflation is measured by the first SR parameter $\epsilon \equiv -\frac{\dot H}{H^2}$
which  is nearly constant and small. During the USR phase the dynamics of the system is given by 
 \ba
\ddot \phi + 3 H \dot \phi =0 \, , \quad \quad
3 M_P^2 H^2 \simeq V_0 \, ,
\ea
where $M_P$ is the reduced Planck mass. 

As the potential is flat during the USR phase, $H$ approaches a fixed value and 
from the field equation we obtain  $\dot \phi \propto a(t)^{-3}$. Correspondingly,   the slow-roll parameter $\epsilon$ falls off exponentially during the USR phase as well, $\epsilon \propto a(t)^{-6}$. On the other hand,  the second slow-roll parameter $\eta \equiv \frac{\dot \epsilon}{H \epsilon} \simeq -6 $  which is the hallmark of the USR phase. 

It is convenient to work with the number of e-fold $N$ as the clock, $d N= H(t) dt$.
We choose the convention that the USR phase starts at $N=0$ so for the first stage of inflation, $N<0$. In particular, the CMB scale modes leave the horizon at around $N\sim -16$. The duration of the USR phase is denoted by $\Delta N$ which is a free parameter of our setup.  Going to conformal time, $d \tau= dt/a(t)$, the 
USR phase is extended during $\tau_i \leq \tau \leq \tau_e$ and the 
duration of USR phase is given by $\Delta N= \ln\big( \frac{\tau_i}{\tau_e}\big)=  \ln\big( \frac{ k_{e}}{ k_{i}}\big)$ in which $k_i (k_e)$ represents the mode which leaves the horizon at the start (end) of USR phase.  The  slow-roll parameter at the end of USR phase $\epsilon_{e}$
is related to its value at the start of USR phase $\epsilon_{i}$ via  $\epsilon_{e}= \epsilon_{i} e^{-6 \Delta N}$.

As explained above, we assume the USR phase is followed by a SR phase. Therefore,  we need to investigate the evolution of the slow-roll parameters $\epsilon(\tau)$ and $\eta(\tau)$ after the USR phase. This was studied in more details in  \cite{Cai:2018dkf}, see also \cite{Cai:2021zsp, Cai:2022erk} for similar studies. Let us denote the final SR parameters with their attractor values 
by $\epsilon_\mathrm V$ and $\eta_\mathrm V$ which are expressed in terms of the first and the second derivatives of the potential in the final SR phase. To simplify the analysis, here, as in \cite{Firouzjahi:2023aum},  we assume that the potential in the final stage is such that    $\epsilon_\mathrm V \gg  |\eta_\mathrm V|$ though this assumption can be relaxed with no important changes in the results. A key parameter in our setup is the sharpness parameter $h$ which controls how quickly the system reaches to its final attractor limit. Following  \cite{Cai:2018dkf}, we define $h$ as follows 
\ba
\label{h-def}
h\equiv \frac{6 \sqrt{2 \epsilon_\mathrm V} }{\dot \phi(t_e)}  = -6 \sqrt{\frac{\epsilon_V}{\epsilon_{{e} }}} \, .
\ea
With this definition, the  slow-roll parameters $\epsilon(\tau)$ and $\eta(\tau)$ after the USR transition are given by 
\ba
\label{ep-eq}
\epsilon(\tau)= \epsilon_ e  \Big[\frac{h}{6} - (1+ \frac{h}{6} ) \big(\frac{\tau}{\tau_e} \big)^3 \Big]^{2}  \,  \quad \quad (\tau > \tau_e) \, ,
\ea
and
\ba
\label{eta-eq}
\eta(\tau) = -\frac{6 (6+h)}{(6+h) - h   \big(\frac{\tau_e}{\tau} \big)^3} \, 
\quad \quad (\tau > \tau_e)  .
\ea
As discussed previously, the above results are obtained in the limit of an instant transition from the USR phase to the final SR phase. Even in this limit it will take some time for the system to reach to its attractor phase which is measured by the sharpness parameter $h$.  In the limit $h \rightarrow -\infty$, the system reaches
its final attractor phase immediately after $\tau_e$ in which the mode functions become frozen. On the other hand, for other values of $h$ the system keeps evolving after $\tau_e$ until $\epsilon(\tau)$ approaches its final  attractor value $\epsilon_\mathrm V$.  A particular case of transition is when $h=-6$ in which $\epsilon(\tau)$ is frozen to its value at the end of USR, $\epsilon(\tau) =\epsilon_e$ with $\eta(\tau)=0$ for $\tau> \tau_e$. 
This  limit was mostly studied in recent literature concerning the loop correction such as in \cite{yokoyama}. In the following analysis, as in \cite{Firouzjahi:2023aum},  we consider a general value of $h$ including the spacial case of $h=-6$. 
Another important point is that the larger the value of  $|h|$ is the larger $\epsilon_V$  is compared to $\epsilon_e$. Correspondingly, the final power spectrum  scales somewhat inversely  with $|h|$. As a result, a larger (smaller) value of $|h|$ yields to a  smaller (larger) final power spectrum. 

We work with the comoving curvature perturbation $\calR$ which in spatially flat gauge is related to inflaton perturbation via $\calR \equiv -\frac{H}{\dot \phi} \delta \phi$. Going to Fourier space, we  define the quantum mode function  in terms of the annihilation and creation operators  as usual via
\ba
\calR(t,  \bfx) = \int \frac{d^3 k}{(2\pi)^3} e^{i {\bf k}\cdot {\bf x}} 
\Big( \calR_k(t) a_{\bf k} + \calR^*_k(t) a_{-\bf k}^\dagger \Big) \, ,
\ea
in which $a_\bfk$ and $a_{\bf k}^\dagger$ satisfy the usual commutation relation associated to the annihilation and creation operators, $[ a_\bfk, a_{\bfk'}^\dagger] = \delta (\bfk -\bfk')$. 

Starting  with the Bunch-Davies (Minkowski) initial condition and imposing the  continuity of $\calR$ and $\dot\calR$ at the transition points $\tau= \tau_{{i}}$ and $\tau= \tau_{{e}}$, the mode function at each stage of  
$SR\rightarrow USR\rightarrow SR$ can be obtained \cite{Firouzjahi:2023aum}. 
The outgoing curvature perturbation  $\calR^{(3)}_{k}(t)$ in the final 
USR phase (third phase) is given by   
 \cite{Firouzjahi:2023aum},
\ba
\label{mode3}
\calR^{(3)}_{k} =  \frac{H}{ M_P\sqrt{4 \epsilon(\tau) k^3}}  
\Big[ \alpha^{(3)}_k ( 1+ i k \tau) e^{- i k \tau}  + \beta^{(3)}_k ( 1- i k \tau) e^{ i k \tau}  \Big] \, , 
\ea
with $\epsilon(\tau)$ given by Eq. (\ref{ep-eq}) and
the coefficients $(\alpha^{(3)}_k, \beta^{(3)}_k)$ are as follow,
\ba
\label{alpha-beta3}
\alpha^{(3)}_k = \frac{1}{8 k^6 \tau_i^3 \tau_e^3}  \Big[ 3h
 ( 1 -i k \tau_e)^2 (1+i k \tau_i)^2 e^{2i k (\tau_e- \tau_i)}
-i (2 k^3 \tau_i^3 + 3i k^2 \tau_i^2 + 3 i) (4 i k^3 \tau_e^3- h k^2 \tau_e^2 - h) \Big]  
\nonumber,
\ea
and
\ba
\beta^{(3)}_k=   \frac{1}{8 k^6 \tau_i^3 \tau_e^3}  \Big[ 3 ( 1+ i k \tau_i)^2 ( h+ h k^2 \tau_e^2 + 4 i k^3 \tau_e^3 ) e^{-2 i k \tau_i} + i h ( 1+ i k \tau_e)^2  ( 3 i + 3 i k^2 \tau_i^2 + 2 k^3 \tau_i^3 ) e^{- 2 i k \tau_e}
 \Big] \nonumber .
\ea

The physical quantities are calculated at the end of inflation $\tau=0$  when the system has reached to its attractor phase with $\epsilon(\tau) \rightarrow \epsilon_V$.  The curvature perturbation power spectrum $\calP_\calR$ from Eq. (\ref{mode3}) is obtained to be
\ba
\label{power-final}
\calP_\calR(k, \tau=0) = \frac{H^2}{ 8 M_P^2 \pi^2  \epsilon_\mathrm V }  
 \big| \alpha^{(3)}_k   + \beta^{(3)}_k \big|^2  \, .
\ea

\begin{figure}[t]
	\hspace{-1.1cm}
	\includegraphics[ width=0.58\linewidth]{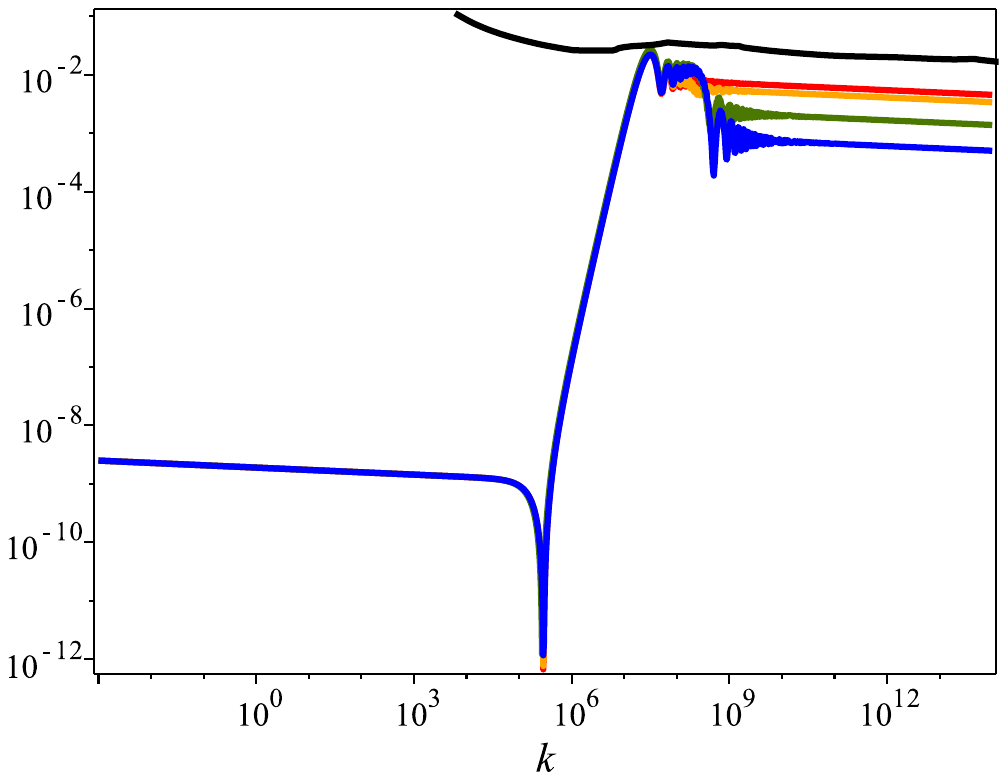}
	\hspace{-0.9cm}
	\includegraphics[ width=0.58\linewidth]{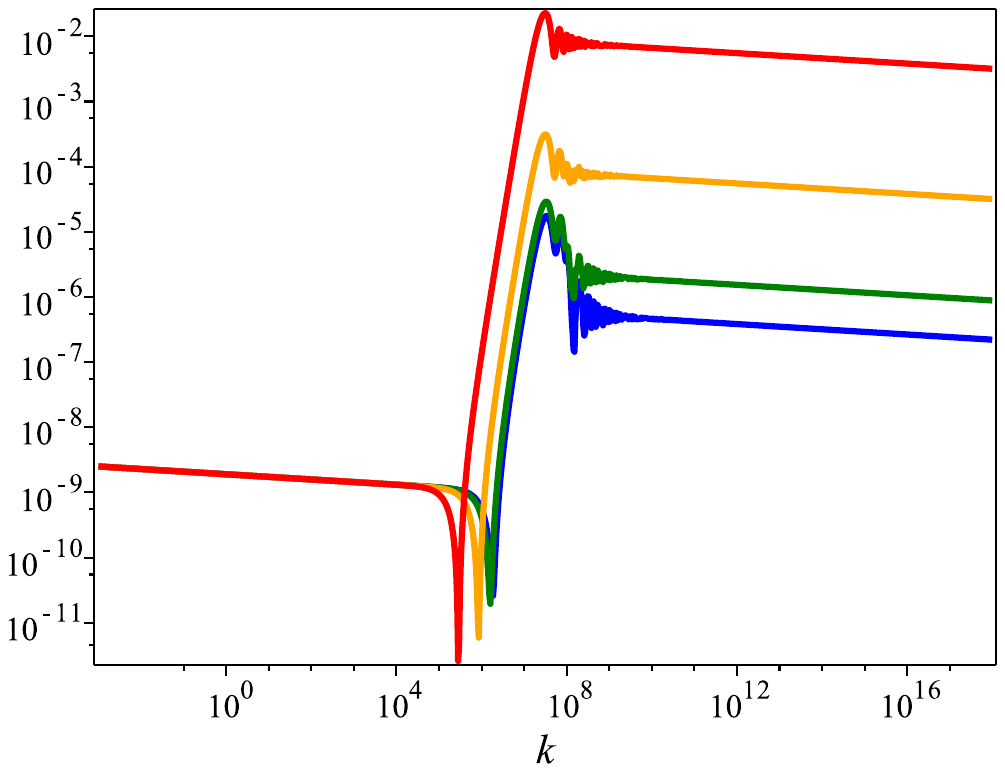}
	\vspace{-6 cm}
	\caption{ The plot of $\calP_\calR$ vs. $k$ for various values of $h$ with $h =-0.1$ (red), $h=-1$ (orange), $h=-6$ (green) and $h=-12$ (blue).  Left: The values of $(h, \Delta N)$ are fixed such that the peak of $\calP_\calR$ does not violate the PBHs bounds shown by the black (top) curve.   All four curves very nearly share the same values of the USR peak and the dip prior to the USR phase.
		Right: We have fixed $\Delta N=1.3$. As $\Delta N$ is fixed for all curves, the power is controlled by $h$ such that the larger $|h|$ is the smaller the final plateau is.  In addition, the position of the dip moves towards the right by increasing $|h|$. 
	}
	\label{plot1}
\end{figure}

The behaviour of  $\calP_\calR(k, \tau=0)$ are plotted in Fig. \ref{plot1}. 
There are a number of common features as can be seen in this figure. 
First, we have the plateau on large scales associated to the  modes which leave the horizon long before the USR phase starts. The amplitude of the power spectrum for 
these perturbations  is fixed by the COBE normalization  on $\mathrm k_{\CMB} = 0.05\,  \mathrm{Mpc}^{-1}$ with  $\calP_\calR \simeq 2.1 \times10^{-9}$.
Second, prior to the USR phase  there is a dip  in power spectrum 
followed  by a universal scaling $\calP \propto k^{4}$. Third, there are oscillations superimposed on the USR plateau after the maximum. Fourth, there is a plateau 
 for the modes which leave the horizon  at the final stage of inflation. 
 As discussed previously, the larger is the value of $|h|$, the smaller is the final power spectrum which we will demonstrate bellow.

Let us look at the parameter dependence of the the power spectrum given in  
Eq. (\ref{power-final}). Two important parameters are  the sharpness of the transition $h$ and the duration of the USR phase $\Delta N$. In addition, we have the energy scale of inflation $H$ and the final slow-roll parameter $\epsilon_V$. As can be seen from  Eq. (\ref{power-final}) the latter two parameters appear in a combination which is fixed by the overall COBE normalization at CMB scales.  
Another independent variable may be considered to be  the starting time of USR phase, $\tau_i$. However, in order to obtain the enhancement in power for PTAs observations, we need the starting time of USR to be when the mode of interest which leaves the horizon has the nano-Hertz frequency. This requires the starting time of USR phase compared to the CMB scales to be separated by about 16 e-folds. Finally, the spectral index $n_s$ is fixed by its best fit value from Planck observation, i.e. $n_s\simeq 0.96$ \cite{Planck:2018vyg, Planck:2018jri}. In summary, at the end we have two main independent parameters: the sharpness parameter $h$ and the duration of USR
$\Delta N$ which we will vary in our analysis below. 

A crucial point is that models with the intermediate USR phase can generate significant PBHs which are constrained by observations. Therefore, in order to meet the various bounds on PBHs formation, we need to impose an additional bound on $(h, \Delta N)$ parameter space. These PBH constraints leave only one of them 
free which we take to be $h$. A view of the power spectrum for various values of $h$ and the bound from PBHs are shown in Fig. \ref{plot1}. 
Schematically, we see that  the PBHs bound is roughly translated into $\calP_\calR < 10^{-2}$.

More precisely, in order to consider the PBHs bound on the curvature power spectrum, we need to know about the probability distribution function (PDF) of the primordial perturbations. In a common approach~\cite{Sasaki:2018dmp}, the mass fraction parameter $\beta$ is related to statistics of ${\cal R}$ as~\cite{Garcia-Bellido:2016dkw}
\begin{align}
\beta \simeq \int_{{\cal R}_c}^{\infty}~f_{\cal R}(x)~ \dd x \simeq \dfrac{1}{2} {\rm Erfc}\left(
\dfrac{{\cal R}_c}{\sqrt{2{\cal P}_{\cal R}}}
\right)
\end{align} 
where $f_{\cal R}$ is PDF of $\cal R$ and ${\cal R}_c \sim {\cal O}(1)$~\cite{Musco:2020jjb,Ozsoy:2023ryl}. The second estimation comes from the fact that we can consider a Gaussian PDF for $\cal R$ with zero mean and variance at the order of power spectrum.

\begin{figure}[t]
	\centering
	\includegraphics[ width=0.8\linewidth]{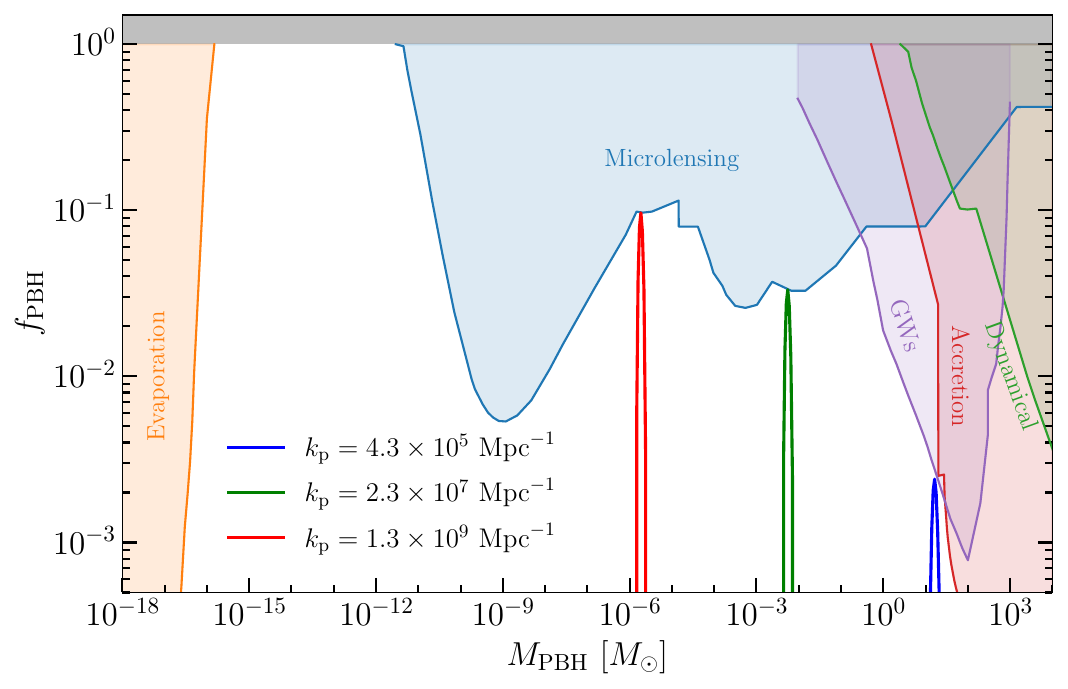}
	\caption{ Fraction $f_{\rm PBH}$ as a function of the mass of the formed
		PBHs in unit of solar mass for USR models in which the power spectrum has a peak around $k_{\rm p}$. 
			The observational
			constraints are taken from Refs. \cite{Green:2020jor, Kavanagh, Carr:2020gox}. The parameters $(h, \Delta N)$ are fixed such that the curve for each $k_{\mathrm{p} }$ reaches the maximum value of power allowed by the PBHs bound. For example, for $k_{\mathrm{p} } =2.3 \times 10^{7}~{\rm Mpc}^{-1}$ (green curve), we have  $(h, \Delta N)=(-0.1, \,  1.45), (-1,  \, 2.17), (-6, \, 2.59)$ and $(-12, \, 2.68)$. These values of $(h, \Delta N)$ are used to generate the results in Figs. \ref{fig: NanoGrav_2} and \ref{fig: NanoGrav_3}.
		}
	\label{fig:fPBH}
\end{figure}

After PBH production, it is crucial to determine the fraction of PBH abundance in dark matter density at the present epoch. It is roughly given by~\cite{Sasaki:2018dmp}
\begin{equation}
\label{fPBH}
f_{\text{PBH}}(M_{\rm PBH}) \simeq 2.7 \times 10^{8} \Big(\dfrac{M_{\rm PBH}}{M_{\odot}}\Big)^{-\frac{1}{2}} \beta(M_{\rm PBH}) \,,
\end{equation} 
where $M_{\odot}$ and $M_{\rm PBH}$ are the solar mass  and the PBH mass respectively. Assuming an instant reheating at the end of inflation~\cite{Ozsoy:2023ryl}, PBH mass can be estimated by
\begin{align}
\dfrac{M_{\rm PBH}}{M_{\odot}} \simeq 30 \, 
\bigg(\dfrac{k_{\rm p}}{3.2 \times 10^5 ~{\rm Mpc}^{-1}}\bigg) ^{-2} \, ,
\end{align}
where $k_{\rm p}$ is the peak scale of perturbations responsible for PBH formation, i.e. the wave-number associated to the maximum of the power spectrum. In Fig.~\ref{fig:fPBH}, we have illustrated the mass function $f_{\rm PBH}$ for various values of $k_{\rm p}$.  PBHs with a given mass have an induced GW counterpart with a peak frequency at around Eq. \eqref{fp}. Therefore,  we have a one-to-one correspondence between the frequency and the PBH mass.

Now  let us look at the various limits of the power spectrum  Eq. (\ref{power-final}). We have two dimensionless numbers, $x\equiv -k \tau_i$ and $e^{-\Delta N}$. First consider the limit  $e^{-\Delta N} \ll x$ so we expand Eq. (\ref{power-final}) to leading order in $e^{-\Delta N}$, obtaining
\begin{align}
\label{power-final2}
\calP_\calR(k, \tau=0) \simeq  &\frac{e^{6 \Delta N}}{2} {\cal P}_\CMB \big(\frac{h-6}{h}\big)^2 
\\  &\times \Big[ 2 x^6 + 9 x^4+ 18 x^2 + 9 
+ (21 x^4 -9) \cos(2 x) 
+ ( 6 x^5 - 24 x^3 - 18 x) \sin(2 x) \Big] \, , \nonumber
\end{align}
in which ${\cal P}_\CMB $ is the CMB scale power spectrum given by 
\ba
{\cal P}_\CMB = \frac{H^2 }{8 \pi^2 M_P^2 \epsilon_i} \, . 
\ea

From  Eq. (\ref{power-final2}) we see that $\calP_\calR \propto e^{6 \Delta N}$ which is the hallmark of USR inflation for the  modes which leave the horizon during the USR phase. Second, we see that  $\calP_\calR \propto \big(\frac{h-6}{h}\big)^2$. This is clearly seen in Fig. \ref{plot1}  as cases with higher value of 
$|h|$  have lower power in the final plateau. 
The physical reason is as follows. Models with larger 
$|h|$ reach the final attractor phase more quickly. For this to happen, $\epsilon(\tau)$ should assume its final value $\epsilon_\mathrm V >  \epsilon_e$ quickly as well. This means that  the mode becomes frozen quickly after the USR phase but with a final  amplitude $\calP_\calR(k, \tau=0) < \calP_\calR(k, \tau_e)$.

To understand the scaling behaviour of the power spectrum prior to USR phase 
and close to the USR peak, let us consider the $x\ll 1$ limit of the expression (\ref{power-final2}), obtaining

\ba
\label{power-small-x}
\calP_\calR(k, \tau=0) \simeq  \frac{2 }{25 } e^{6 \Delta N} {\cal P}_\CMB \big(\frac{h-6}{h}\big)^2 \, (k \tau_i)^4  \, .
\ea
It shows that  the power spectrum scales like  $\calP_\calR(k) \propto k^4$ prior to and after the USR phase starts, a phenomenon which was observed  previously in \cite{Byrnes:2018txb, Cole:2022xqc, Carrilho:2019oqg, Ozsoy:2021pws, Pi:2022zxs} as well. 

As we see in  Fig. \ref{plot1},  there is a dip in power spectrum prior to USR phase where the above mentioned $k^4$ scaling starts. To understand the nature of this dip, note that the expression  (\ref{power-final2}) is obtained assuming that $e^{-\Delta N} \ll x$. However, this limit is violated for very long modes  which become superhorizon much earlier than the USR phase starts. In particular, the CMB scale modes belong to this limit. Considering the $x \ll e^{-\Delta N}$   limit of the power spectrum we obtain
\ba
\label{power-small-x2}
\calP_\calR(k, \tau=0) \simeq  {\cal P}_\CMB \Big( 1- \frac{4}{5} \frac{h-6}{h} 
\, (k \tau_i)^2 
\Big) \, , \quad \quad   (k \tau_i \rightarrow 0 ) \, .
\ea 
The position of the dip $k= k_\mathrm d$ is  where the two expressions 
(\ref{power-small-x2}) and (\ref{power-small-x}) become comparable, 
yielding to the approximate value (see also  \cite{Pi:2022zxs}) 
\ba
\label{d-dip}
k_\mathrm d  \tau_i\simeq   \sqrt{\frac{5 h}{4(h-6)}}\,  e^{-\frac{3}{2} \Delta N} \, .
\ea
From the above formula we see that for a fixed value of $\Delta N$, 
as  $|h|$ increase the value of $k_\mathrm d$ increases as well, i.e. the dip moves towards the right, as seen in the right panel of  
Fig. \ref{plot1}.

As we mentioned previously  the USR model can generate non-Gaussianities. However, the magnitude of $f_{NL}$ depends on $k$ as well. For the mode which leaves the horizon during the early stage of inflation and prior to USR phase, then the Maldacena consistency condition does hold and for these modes $f_{NL}$ is basically very small. On the other hand, for the modes which leave the horizon during the USR phase, i.e. for $k_i< k< k_e$, the consistency condition is violated. The final value of
$f_{NL}$ for these modes crucially depends on the parameter $h$. This was studied in details in \cite{Cai:2018dkf} and \cite{Firouzjahi:2023aum} in which it is shown that up to slow-roll corrections, 
\ba
\label{fNL}
f_{NL} = \frac{5 h^2}{2 (h-6)^2} \, .
\ea
For an infinitely sharp transition with $h\rightarrow -\infty$ in which the system assumes its final attractor phase immediately after the USR transition, we obtain the maximum value $f_{NL}=\frac{5}{2}$. However, lowering $|h|$ reduces $f_{NL}$ accordingly.
For example, for the standard scenario in which $h=-6$ as studied in \cite{yokoyama} one obtains $f_{NL}=\frac{5}{8}\simeq 0.63$. For milder transitions with  $  |h| \lesssim 1$, from Eq. (\ref{fNL}) we typically obtain $f_{NL} \ll 1$. For example for $h=-1$ and $h=-0.1$ which we will study below, we obtain $f_{NL} \simeq 0.051$ and $f_{NL} \simeq 0.0007$ respectively. Therefore, to very good approximation one can employ  the Gaussian bound on PBH's formation. To be more specific, to neglect the non-Gaussianity effects in PBH formation, we need that $f_{NL} \calP_\calR \ll 1$ \cite{Young:2013oia}. In our model with the maximum value  $f_{NL}= \frac{5}{2}$, we can easily satisfy the Gaussianity approximation if $\calP_\calR$ is small. In our analysis, as can be seen in  Fig. \ref{plot1}, the PBHs bound typically require that $\calP_\calR \lesssim 10^{-2}$ so we easily meet the condition  $f_{NL} \calP_\calR \ll 1$ for all practical values of $h$.

As mentioned in Introduction section, the loop correction is an open question 
in this setup \cite{yokoyama, Kristiano:2023scm, Riotto:2023hoz, Riotto:2023gpm, Choudhury:2023jlt, Choudhury:2023rks, Firouzjahi:2023aum, hr, Motohashi:2023syh,  f, t, Firouzjahi:2023btw, Fumagalli:2023hpa}. The effects of the sharpness of the transition on the loop corrections were studied in \cite{Firouzjahi:2023aum}. It was shown that for an arbitrarily  sharp transition the one-loop corrections become very large. More specifically, it was shown in \cite{Firouzjahi:2023aum} that for $|h| \gg 1$, the one-loop corrections scale linearly with $h$, yielding to a large loop correction in line with the results of  \cite{yokoyama, Kristiano:2023scm}. However, it was shown in \cite{hr} that for a mild transition the one-loop corrections to CMB scale modes are slow-roll suppressed and are harmless.
In our current study, in order to trust the Gaussian approximation for PBHs formation and to neglect the large loop corrections, one requires $| h| \lesssim 1$ 
which we shall assume in our analysis.
However, in order to compare the predictions of the setup with both sharp and mild transitions, we also present the results for SIGWs for the cases of sharp transition as well. In our numerical plots below, the examples of sharp transitions correspond to the cases $h=-6$ and $h=-12$. As another perspective to the question of loop corrections,   it was argued in \cite{Fumagalli:2023hpa} that by considering the surface terms the leading loop corrections will disappear. This was studied for the cubic Hamiltonian  in \cite{Fumagalli:2023hpa} while the contribution associated to quartic Hamiltonian and the surface terms is an open question.

\section{SIGWs and Constraints from PTAs Observations} 

The curvature perturbations, which have been generated during inflation, can re-enter the horizon during the radiation-dominated (RD) era in which the metric (in conformal Newtonian gauge) reads 
\begin{align}
\dd s^2 =& -a^2\left[ (1+2\Phi) \dd \tau^2 + \Big((1 - 2 \Psi)\delta_{ij} +\dfrac{1}{2}h_{ij} \Big) \dd x^i \dd x^j\right]. 
\label{eq:metric}
\end{align}
Here $\tau$ is the conformal time during RD era, $\Phi$ and $\Psi$ are the Bardeen potentials and $h_{ij}$ is the transverse-traceless tensor perturbations. Using the Einstein's field equations, the evolution of Fourier modes of $h_{ij}$, denoted by $h^{\lambda}_{\bf{k}}$, are given by
\begin{align}
{h^{\lambda}_{\bf{k}}}''\!(\eta) + 2 \mathcal H {h^{\lambda}_{\bf{k}}}'\!(\eta) + k^2 h^{\lambda}_{\bf{k}}(\eta) = 4 S^{\lambda}_{\bf{k}}(\eta),
\label{eq:eom_h}
\end{align}
where $\lambda$ represents  the two polarizations.  The primes denote the derivative with respect to the conformal time, $\mathcal H = a'/a$ is the conformal Hubble parameter, and the source term $S^\lambda_{\bf k}$ is transverse and  traceless which is second order in scalar perturbations, given by 
\begin{align}
\label{eq:s}
S^\lambda_{\bf{k}} =&\,  \int \! \frac{\dd^3 q}{(2\pi)^3} \, \varepsilon^{\lambda}_{ij}(\hat{\bf k})~q^i q^j \bigg[ 2 \Phi_{\bf{q}} \Phi_{\bf{k-q}}  
+   \left(\mathcal{H}^{-1}\Phi'_{\bf{q}} + \Phi_{\bf{q}} \right)
\left( \mathcal{H}^{-1}\Phi'_{\bf{k-q}} + \Phi_{\bf{k-q}}\right) \bigg] \, ,
\end{align}
where $\varepsilon^{\lambda}_{ij}$ is the polarization tensor. Note that here we have neglected the vector perturbations and the anisotropic stress ($\Phi \simeq \Psi$). 

In Fourier space, the Bardeen potential is related to ${\cal R}_{\bf k}$ through transfer function ${\cal T}({\bf k}\tau)$ as
\begin{align}
\Phi_{\bf k} = \dfrac{2}{3}{\cal T}({\bf k}\tau){\cal R}_{\bf k} \, .
\end{align}
The transfer function encodes the linear evolution of the Newtonian potential after horizon reentry which has a  oscillatory behaviour. Solving the equation of motion \eqref{eq:eom_h}, taking the late-time limit during a RD era ($\tau \rightarrow \infty$ at the matter-radiation equality), the power spectrum of tensor fluctuations is given by \cite{Kohri:2018awv}
\begin{align}
\label{eq:P_h}
{\cal P}_h(\tau ,k) = 4\int_0^\infty \dd v \int_{|1-v|}^{|1+v|} \dd u~ ~\mathcal{K}\left(u,v,k \tau \right)~ \mathcal{P}_\mathcal{\cal R}\left(k u\right) \mathcal{P}_\mathcal{\cal R}\left(k v\right) \,,
\end{align} 
For further details about the integration kernel $\mathcal{K}$ and how to perform the integrations see \cite{Kohri:2018awv}. 

The produced GW contributes to the total energy of Universe and dilutes like radiation. Taking into account the following matter-dominated and dark-energy-dominated eras, the current value of $\Omega_{\rm GW}$, the fraction of the GW energy density per logarithmic wavelength, is obtained to be
\begin{align}
\label{eq: Omega_GW}
\Omega_{\rm GW}h_0^2 =\Omega_{\rm r}h_0^2~ \left(\dfrac{g_*}{g_{*,e}}\right)^{1/3}\Omega_{{\rm GW},e}(f) \,.
\end{align}
Here $\Omega_{\rm r}$ is the present-day
abundance of radiation, $g_*$ is the number of relativistic degrees of freedom in energy density, and  the subscripts $e$ denotes the time of emission. Note that $\Omega_{\rm r}h_0^2 \simeq 4.2 \times 10^{-5}$ with $h_0=H_0 /100 \,{\rm km} \,{\rm s}^{-1}\,{\rm Mpc}^{-1}$. Here $f=c \, k/(2\pi)$ is the frequency of the GW which has appeared due to the $k$-dependence of the spectrum of curvature perturbations \eqref{power-final} during inflation.

\begin{figure}[t]
	\centering
	\includegraphics[ width=0.68\linewidth]{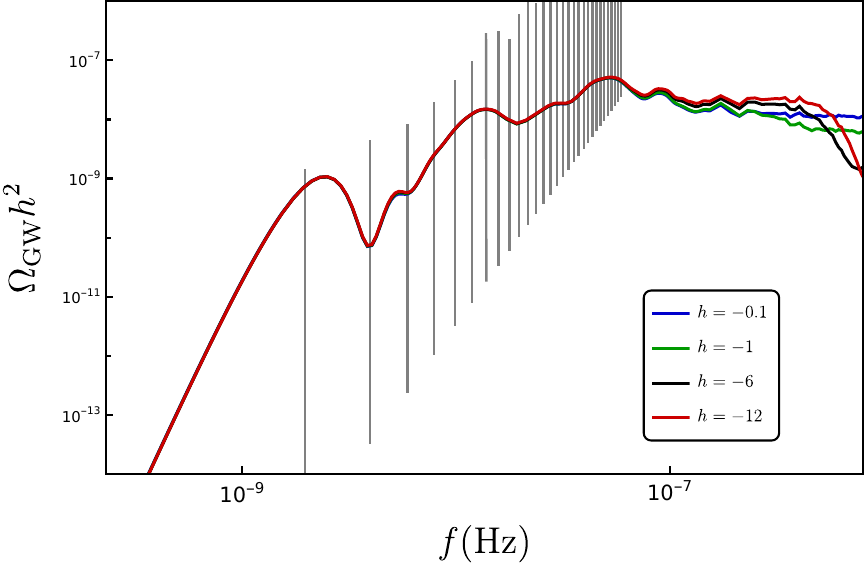}
	\caption{ The prediction of the model compared to the NANOGrav data \cite{NANOGrav:2023hvm} for a power spectrum with the peak scale $k_{\mathrm{p} } \simeq  10^{7}~{\rm Mpc}^{-1}$ but for various values of $h$. The  value of $\Delta N$ is fixed by the PBHs fraction allowed  in Fig. \ref{fig:fPBH}.  More specifically, $(h, \Delta N) = (-0.1, \, 1.45), (-1, \, 2.17) , (-6,\, 2.59)$ and $ (-12,\, 2.68)$. }
	\label{fig: NanoGrav_2}
\end{figure}

We have used the curvature perturbations power spectrum  \eqref{power-final} generating during the USR phase and calculated the convolution integral \eqref{eq:P_h} numerically to find the current fractional energy density of GWs \eqref{eq: Omega_GW}. The results are shown in Fig.~\ref{fig: NanoGrav_2} for various values of $h$ and $\Delta N$ in nano-Hertz bound. The results have been presented alongside the posteriors of an Hellings-Downs (HD) correlated free spectral reconstruction of the NANOGrav signal \cite{NANOGrav:2023hvm}. The values of $(h, \Delta N)$ are fixed such that the peak of ${\cal P}_{\cal R}$ does not violate the PBHs bounds. As seen, the spectra of our model follow the HD pattern expected for a stochastic gravitational wave background. Interestingly, we see that within the observable window the predictions of all models, mild ($|h| \lesssim 1$) or sharp ($|h | \gg 1$), follow the same pattern and are not significantly different from each other. However, outside the NANOGrav observed window ($f > 10^2 \, \mathrm{nHz}$) the curves deviate from each other noticeably. This pattern is similar to the plots of
power spectrum of curvature perturbations presented in Fig. \ref{plot1}. The reason is that all curves are obtained after imposing the PBHs bounds. 
However, the starting time of the USR and  the value of the peak of the USR plateau are very similar for all curves as seen in Fig. \ref{plot1}. This is the reason why all curves, sharp or mild, follow close trajectories on the observable window. 

\begin{figure}[t]
	\centering
	\includegraphics[ width=0.85\linewidth]{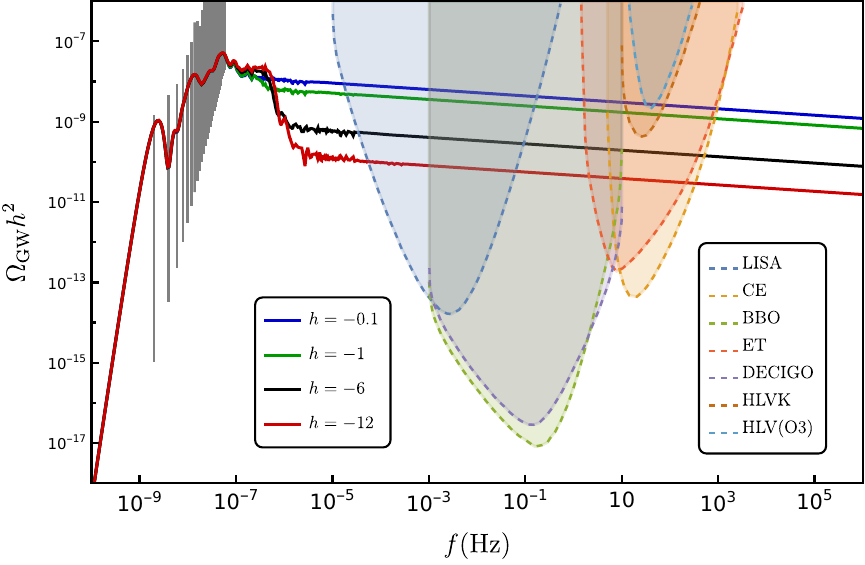}
	\caption{ The same as in Fig. \ref{fig: NanoGrav_2} but for an extended range of the frequency. As we see, depending on the value of $h$, the tail of SIGWs differ noticeably for each model which can be tested in future observations. 
The higher is the value of $|h|$ the lower is the value of the SIGW tail.
	}
	\label{fig: NanoGrav_3}
\end{figure}

However, crucial to our setup is that outside the  NANOGrav window, the curves have distinct predictions for SIGWs on frequencies much higher than  
$\sim 10^2 \, \mathrm{nHz}$.  More specifically, the final tail of the power spectrum scales somewhat inversely with the sharpness parameter $h$ such that milder (sharper) transitions have higher (lower) tails. In Fig.~\ref{fig: NanoGrav_3} we  we have shown the SIGWs spectra for a larger frequency range. In this figure, the quantity $\Omega_{\rm GW} h^2$ was plotted against the frequency together with the sensitivity of the various existing and forthcoming GW experiments such as LISA, SKA, BBO, DECIGO etc. As seen, the tails of SIGW spectrums for different $(h, \Delta N)$ can fall into the sensitivity bounds  of these observations. It means that different values of $(h, \Delta N) $ are distinguishable from each other in future GWs observations.

\section{Summary and Discussions}

 The stochastic gravitational wave background detected by various PTAs observations can open a new window to the dynamics of early universe. In particular, this signal  can be generated by GWs induced by scalar perturbations at second order in perturbation theory. The SIGWs can be used as a tool to distinguish various inflationary scenarios. A key criteria is that the models which are employed to explain the SGWB observed by the PTAs observations should not generate too much of PBHs which are constrained in various frequency ranges.
 
 In this work we have considered a single field model of inflation containing an intermediate USR phase. This setup has been used extensively in the past to generate PBHs and for the induced GWs studies. We have paid particular attention to the sharpness parameter of the model which play significant roles in loop corrections and for the amplitude of non-Gaussianity. In order to be away from the dangerous one-loop corrections we require a mild transition with $| h| \lesssim 1$. This is also the limit where the amplitude of non-Gaussianity is small and one can employ the Gaussian predictions of the PBHs formation. 
 
 We have calculated the spectrum of SIGWs and compared it to the NANOGrave results. The predictions of the model are consistent with the observed data.  However, a careful data analysis is required to contrast the predictions of the model with the PTA datas and to  put constraints on the model parameters.
 
 While our setup can qualitatively explain the origin of the NANOGrave observations, but it has specific predictions for the spectrum in higher frequency ranges. Our model predicts that a setup with a mild (sharp) transition has a higher (lower) tail of IGWs once they are fit to the current NANOGrave data.   As a result, the predictions of our model for the amplitude of induced GWs can be tested in future GWs observation which may put constraints on the model parameter or to rule it out.

\vspace{0.9cm}
   
 {\bf Acknowledgments:}  We are grateful to   Anotonio Riotto, Sina Hooshangi, Seyed Ali Hosseini Mansoori, and Ali Akbar Abolhasani for useful  discussions and comments on the draft.  We are  partially supported by the ``Saramadan" Federation of Iran. A. T. would like to thank University of Rwanda, EAIFR, and ICTP  for their kind hospitalities during the 17th international workshop on the ``Dark Side of the Universe" where this work was in its final stages.


\begin{thebibliography}{99}



\bibitem{NANOGrav:2023gor}
G.~Agazie \textit{et al.} [NANOGrav],
Astrophys. J. Lett. \textbf{951}, no.1, L8 (2023)
doi:10.3847/2041-8213/acdac6
[arXiv:2306.16213 [astro-ph.HE]].

\bibitem{Reardon:2023gzh}
D.~J.~Reardon \textit{et al,} 
Astrophys. J. Lett. \textbf{951}, no.1, L6 (2023)
doi:10.3847/2041-8213/acdd02
[arXiv:2306.16215 [astro-ph.HE]].

\bibitem{Antoniadis:2023ott}
J.~Antoniadis \textit{et al,}  
[arXiv:2306.16214 [astro-ph.HE]].

\bibitem{Xu:2023wog}
H.~Xu \textit{et al,}  
doi:10.1088/1674-4527/acdfa5
[arXiv:2306.16216 [astro-ph.HE]].



\bibitem{Kosowsky:1992rz}
A.~Kosowsky, M.~S.~Turner and R.~Watkins,
Phys. Rev. Lett. \textbf{69}, 2026-2029 (1992).

\bibitem{Kamionkowski:1993fg}
M.~Kamionkowski, A.~Kosowsky and M.~S.~Turner,
Phys. Rev. D \textbf{49}, 2837-2851 (1994).

\bibitem{Caprini:2007xq}
C.~Caprini, R.~Durrer and G.~Servant,
Phys. Rev. D \textbf{77}, 124015 (2008).

\bibitem{Hindmarsh:2013xza}
M.~Hindmarsh, S.~J.~Huber, K.~Rummukainen and D.~J.~Weir,
Phys. Rev. Lett. \textbf{112}, 041301 (2014).

\bibitem{Kibble:1976sj}
T.~W.~B.~Kibble,
J. Phys. A \textbf{9}, 1387-1398 (1976)
doi:10.1088/0305-4470/9/8/029

\bibitem{Vilenkin:1981bx}
A.~Vilenkin,
Phys. Lett. B \textbf{107}, 47-50 (1981)
doi:10.1016/0370-2693(81)91144-8


\bibitem{Caldwell:1991jj}
R.~R.~Caldwell and B.~Allen,
Phys. Rev. D \textbf{45}, 3447-3468 (1992)
doi:10.1103/PhysRevD.45.3447

\bibitem{Vilenkin:1981zs}
A.~Vilenkin,
Phys. Rev. D \textbf{23}, 852-857 (1981)
doi:10.1103/PhysRevD.23.852


\bibitem{NANOGrav:2023hvm}
A.~Afzal \textit{et al.} [NANOGrav],
Astrophys. J. Lett. \textbf{951}, no.1, L11 (2023).

\bibitem{Antoniadis:2023zhi}
J.~Antoniadis \textit{et al,} 
[arXiv:2306.16227 [astro-ph.CO]].


\bibitem{NANOGrav:2020bcs}
Z.~Arzoumanian \textit{et al.} [NANOGrav],
Astrophys. J. Lett. \textbf{905}, no.2, L34 (2020). 

\bibitem{Vagnozzi:2020gtf}
S.~Vagnozzi,
Mon. Not. Roy. Astron. Soc. \textbf{502}, no.1, L11-L15 (2021).

\bibitem{Zhou:2020kkf}
Z.~Zhou, J.~Jiang, Y.~F.~Cai, M.~Sasaki and S.~Pi,
Phys. Rev. D \textbf{102}, no.10, 103527 (2020)



\bibitem{Domenech:2020ers}
G.~Dom\`enech and S.~Pi,
Sci. China Phys. Mech. Astron. \textbf{65}, no.3, 230411 (2022)

\bibitem{Dandoy:2023jot}
V.~Dandoy, V.~Domcke and F.~Rompineve,
[arXiv:2302.07901 [astro-ph.CO]].


\bibitem{Ananda:2006af}
K.~N.~Ananda, C.~Clarkson and D.~Wands,
Phys. Rev. D \textbf{75}, 123518 (2007). 


\bibitem{Baumann:2007zm}
D.~Baumann, P.~J.~Steinhardt, K.~Takahashi and K.~Ichiki,
Phys. Rev. D \textbf{76}, 084019 (2007). 

\bibitem{Bugaev:2009zh}
E.~Bugaev and P.~Klimai,
Phys. Rev. D \textbf{81}, 023517 (2010). 


\bibitem{Assadullahi:2009nf}
H.~Assadullahi and D.~Wands,
Phys. Rev. D \textbf{79}, 083511 (2009). 

\bibitem{Alabidi:2012ex}
L.~Alabidi, K.~Kohri, M.~Sasaki and Y.~Sendouda,
JCAP \textbf{09}, 017 (2012). 

\bibitem{Cai:2018dig}
R.~g.~Cai, S.~Pi and M.~Sasaki,
Phys. Rev. Lett. \textbf{122}, no.20, 201101 (2019). 

\bibitem{Pi:2020otn}
S.~Pi and M.~Sasaki,
JCAP \textbf{09}, 037 (2020). 


\bibitem{Balaji:2022dbi}
S.~Balaji, G.~Domenech and J.~Silk,
JCAP \textbf{09}, 016 (2022). 

\bibitem{Talebian:2022cwk}
A.~Talebian, S.~A.~Hosseini Mansoori and H.~Firouzjahi,
Astrophys. J. \textbf{948}, no.1, 48 (2023)


\bibitem{Domenech:2021ztg}
G.~Dom\`enech,
Universe \textbf{7}, no.11, 398 (2021). 



\bibitem{Carr:2016drx}
B.~Carr, F.~Kuhnel and M.~Sandstad,
Phys. Rev. D \textbf{94}, no.8, 083504 (2016).

\bibitem{Carr:2020xqk}
B.~Carr and F.~Kuhnel,
Ann. Rev. Nucl. Part. Sci. \textbf{70}, 355-394 (2020).

\bibitem{Sasaki:2018dmp}
M.~Sasaki, T.~Suyama, T.~Tanaka and S.~Yokoyama,
Class. Quant. Grav. \textbf{35}, no.6, 063001 (2018). 


\bibitem{Ozsoy:2023ryl}
O.~\"Ozsoy and G.~Tasinato,
[arXiv:2301.03600 [astro-ph.CO]].

\bibitem{Byrnes:2021jka}
C.~T.~Byrnes and P.~S.~Cole,
[arXiv:2112.05716 [astro-ph.CO]].

\bibitem{Ivanov:1994pa}
P.~Ivanov, P.~Naselsky and I.~Novikov,
Phys. Rev. D \textbf{50}, 7173-7178 (1994). 

\bibitem{Garcia-Bellido:2017mdw}
J.~Garcia-Bellido and E.~Ruiz Morales,
Phys. Dark Univ. \textbf{18}, 47-54 (2017). 

\bibitem{Biagetti:2018pjj}
M.~Biagetti, G.~Franciolini, A.~Kehagias and A.~Riotto,
JCAP \textbf{07}, 032 (2018). 

\bibitem{Ragavendra:2020sop}
H.~V.~Ragavendra, P.~Saha, L.~Sriramkumar and J.~Silk,
Phys. Rev. D \textbf{103}, no.8, 083510 (2021). 

\bibitem{Di:2017ndc}
H.~Di and Y.~Gong,
JCAP \textbf{07}, 007 (2018). 

\bibitem{Liu:2020oqe}
J.~Liu, Z.~K.~Guo and R.~G.~Cai,
Phys. Rev. D \textbf{101}, no.8, 083535 (2020)

\bibitem{Hooshangi:2022lao}
S.~Hooshangi, A.~Talebian, M.~H.~Namjoo and H.~Firouzjahi,
Phys. Rev. D \textbf{105}, no.8, 083525 (2022)

\bibitem{Hooshangi:2023kss}
S.~Hooshangi, M.~H.~Namjoo and M.~Noorbala,
[arXiv:2305.19257 [astro-ph.CO]].

\bibitem{Ghoshal:2023wri}
A.~Ghoshal, A.~Moursy and Q.~Shafi,
[arXiv:2306.04002 [hep-ph]].

\bibitem{Kinney:2005vj} 
  W.~H.~Kinney,
  Phys.\ Rev.\ D {\bf 72}, 023515 (2005)
  [gr-qc/0503017].
  
\bibitem{Morse:2018kda}
M.~J.~P.~Morse and W.~H.~Kinney,
Phys. Rev. D \textbf{97}, no.12, 123519 (2018). 

\bibitem{Lin:2019fcz}
W.~C.~Lin, M.~J.~P.~Morse and W.~H.~Kinney,
JCAP \textbf{09}, 063 (2019). 

  
\bibitem{Namjoo:2012aa} 
  M.~H.~Namjoo, H.~Firouzjahi and M.~Sasaki,
  Europhys.\ Lett.\  {\bf 101}, 39001 (2013).

\bibitem{Maldacena:2002vr} 
  J.~M.~Maldacena,
  JHEP {\bf 0305}, 013 (2003)
  [astro-ph/0210603].

\bibitem{Creminelli:2004yq}
P.~Creminelli and M.~Zaldarriaga,
JCAP \textbf{10}, 006 (2004). 


\bibitem{Martin:2012pe}
J.~Martin, H.~Motohashi and T.~Suyama,
Phys. Rev. D \textbf{87}, no.2, 023514 (2013). 
  
  
\bibitem{Chen:2013aj} 
  X.~Chen, H.~Firouzjahi, M.~H.~Namjoo and M.~Sasaki,
  Europhys.\ Lett.\  {\bf 102}, 59001 (2013). 
 
\bibitem{Chen:2013eea} 
  X.~Chen, H.~Firouzjahi, E.~Komatsu, M.~H.~Namjoo and M.~Sasaki,
  JCAP {\bf 1312}, 039 (2013). 
  
\bibitem{Akhshik:2015rwa}
M.~Akhshik, H.~Firouzjahi and S.~Jazayeri,
JCAP \textbf{12}, 027 (2015). 

\bibitem{Mooij:2015yka}
S.~Mooij and G.~A.~Palma,
JCAP \textbf{11}, 025 (2015). 

\bibitem{Bravo:2017wyw}
R.~Bravo, S.~Mooij, G.~A.~Palma and B.~Pradenas,
JCAP \textbf{05}, 024 (2018). 

\bibitem{Finelli:2017fml}
B.~Finelli, G.~Goon, E.~Pajer and L.~Santoni,
Phys. Rev. D \textbf{97}, no.6, 063531 (2018). 

\bibitem{Pi:2022ysn}
S.~Pi and M.~Sasaki,
Phys. Rev. Lett. \textbf{131}, no.1, 011002 (2023). 

\bibitem{Cai:2018dkf}
Y.~F.~Cai, X.~Chen, M.~H.~Namjoo, M.~Sasaki, D.~G.~Wang and Z.~Wang,
JCAP \textbf{05}, 012 (2018). 



\bibitem{yokoyama}
J.~Kristiano and J.~Yokoyama,
 arXiv{2211.03395}{hep-th}.
 
\bibitem{Kristiano:2023scm}
J.~Kristiano and J.~Yokoyama,
[arXiv:2303.00341 [hep-th]].

\bibitem{Riotto:2023hoz}
A.~Riotto,
[arXiv:2301.00599 [astro-ph.CO]].

\bibitem{Riotto:2023gpm}
A.~Riotto,
[arXiv:2303.01727 [astro-ph.CO]].

\bibitem{Choudhury:2023jlt}
S.~Choudhury, S.~Panda and M.~Sami,
[arXiv:2302.05655 [astro-ph.CO]].

\bibitem{Choudhury:2023rks}
S.~Choudhury, S.~Panda and M.~Sami,
[arXiv:2303.06066 [astro-ph.CO]].



\bibitem{Firouzjahi:2023aum}
H.~Firouzjahi,
[arXiv:2303.12025 [astro-ph.CO]].

\bibitem{hr}
H.~Firouzjahi and A.~Riotto,
arXiv{2304.07801}{astro-ph.CO}.

\bibitem{Motohashi:2023syh}
H.~Motohashi and Y.~Tada,
[arXiv:2303.16035 [astro-ph.CO]].

\bibitem{f}
G.~Franciolini, A.~Iovino, Junior., M.~Taoso and A.~Urbano,
arXiv{2305.03491}{astro-ph.CO}.

\bibitem{t}
G.~Tasinato,
arXiv{2305.11568}{hep-th}.


\bibitem{Firouzjahi:2023btw}
H.~Firouzjahi,
[arXiv:2305.01527 [astro-ph.CO]].

\bibitem{Fumagalli:2023hpa}
J.~Fumagalli,
[arXiv:2305.19263 [astro-ph.CO]].


\bibitem{Wang:2023ost}
S.~Wang, Z.~C.~Zhao, J.~P.~Li and Q.~H.~Zhu,
[arXiv:2307.00572 [astro-ph.CO]].

\bibitem{Vagnozzi:2023lwo}
S.~Vagnozzi,
[arXiv:2306.16912 [astro-ph.CO]].


\bibitem{Franciolini:2023pbf}
G.~Franciolini, A.~Iovino, Junior., V.~Vaskonen and H.~Veermae,
[arXiv:2306.17149 [astro-ph.CO]].

\bibitem{Cai:2023dls}
Y.~F.~Cai, X.~C.~He, X.~Ma, S.~F.~Yan and G.~W.~Yuan,
[arXiv:2306.17822 [gr-qc]].

\bibitem{Inomata:2023zup}
K.~Inomata, K.~Kohri and T.~Terada,
[arXiv:2306.17834 [astro-ph.CO]].

\bibitem{Liu:2023ymk}
L.~Liu, Z.~C.~Chen and Q.~G.~Huang,
[arXiv:2307.01102 [astro-ph.CO]].

\bibitem{Yi:2023mbm}
Z.~Yi, Q.~Gao, Y.~Gong, Y.~Wang and F.~Zhang,
[arXiv:2307.02467 [gr-qc]].

\bibitem{Figueroa:2023zhu}
D.~G.~Figueroa, M.~Pieroni, A.~Ricciardone and P.~Simakachorn,
[arXiv:2307.02399 [astro-ph.CO]].

\bibitem{Gu:2023mmd}
B.~M.~Gu, F.~W.~Shu and K.~Yang,
[arXiv:2307.00510 [astro-ph.CO]].

\bibitem{Ebadi:2023xhq}
R.~Ebadi, S.~Kumar, A.~McCune, H.~Tai and L.~T.~Wang,
[arXiv:2307.01248 [astro-ph.CO]].

\bibitem{Madge:2023cak}
E.~Madge, E.~Morgante, C.~P.~Ib\'a\~nez, N.~Ramberg and S.~Schenk,
[arXiv:2306.14856 [hep-ph]].





\bibitem{Cai:2021zsp}
Y.~F.~Cai, X.~H.~Ma, M.~Sasaki, D.~G.~Wang and Z.~Zhou,
Phys. Lett. B \textbf{834}, 137461 (2022). 

\bibitem{Cai:2022erk}
Y.~F.~Cai, X.~H.~Ma, M.~Sasaki, D.~G.~Wang and Z.~Zhou,
JCAP \textbf{12}, 034 (2022). 


\bibitem{Planck:2018vyg}
N.~Aghanim \textit{et al.} [Planck],
Astron. Astrophys. \textbf{641}, A6 (2020)
[erratum: Astron. Astrophys. \textbf{652}, C4 (2021)].


\bibitem{Planck:2018jri}
Y.~Akrami \textit{et al.} [Planck],
Astron. Astrophys. \textbf{641}, A10 (2020).


\bibitem{Garcia-Bellido:2016dkw}
J.~Garcia-Bellido, M.~Peloso and C.~Unal,
JCAP \textbf{12}, 031 (2016).

\bibitem{Musco:2020jjb}
I.~Musco, V.~De Luca, G.~Franciolini and A.~Riotto,
Phys. Rev. D \textbf{103}, no.6, 063538 (2021)


\bibitem{Byrnes:2018txb}
C.~T.~Byrnes, P.~S.~Cole and S.~P.~Patil,
JCAP \textbf{06}, 028 (2019), 
[arXiv:1811.11158 [astro-ph.CO]].

\bibitem{Cole:2022xqc}
P.~S.~Cole, A.~D.~Gow, C.~T.~Byrnes and S.~P.~Patil,
[arXiv:2204.07573 [astro-ph.CO]].

\bibitem{Carrilho:2019oqg}
P.~Carrilho, K.~A.~Malik and D.~J.~Mulryne,
Phys. Rev. D \textbf{100}, no.10, 103529 (2019), 
[arXiv:1907.05237 [astro-ph.CO]].

\bibitem{Ozsoy:2021pws}
O.~\"Ozsoy and G.~Tasinato,
Phys. Rev. D \textbf{105}, no.2, 023524 (2022).

\bibitem{Pi:2022zxs}
S.~Pi and J.~Wang,
JCAP \textbf{06}, 018 (2023). 

\bibitem{Young:2013oia}
S.~Young and C.~T.~Byrnes,
JCAP \textbf{08}, 052 (2013).


\bibitem{Green:2020jor}
A.~M.~Green and B.~J.~Kavanagh,
J. Phys. G \textbf{48}, no.4, 043001 (2021)

\bibitem{Kavanagh}
Kavanagh, B. J. 2019, bradkav/PBHbounds: Release
version, 1.0, Zenodo, doi: 10.5281/zenodo.3538999. 

\bibitem{Carr:2020gox}
B.~Carr, K.~Kohri, Y.~Sendouda and J.~Yokoyama,
Rept. Prog. Phys. \textbf{84}, no.11, 116902 (2021). 


\bibitem{Kohri:2018awv}
K.~Kohri and T.~Terada,
Phys. Rev. D \textbf{97}, no.12, 123532 (2018). 

\end{thebibliography}


\end{document}